\documentclass{article}
\usepackage{spconf,amsmath,graphicx,hyperref}
\usepackage{bm}
\usepackage{multirow,multicol}
\usepackage{amsmath}
\usepackage{amssymb}
\usepackage{graphicx}
\usepackage{cite}
\usepackage{epsf}
\usepackage{times}
\usepackage{color}
\usepackage{tikz}
\usepackage{xcolor}
\usepackage[ruled,vlined]{algorithm2e}

\title{A Directional-Derivative-Constrained Method for Continuously Steerable Differential Beamformers with Uniform Circular Arrays}
\name{\begin{tabular}{c}
		Tiantian Xiong$^1$, Yongyi Deng$^1$, Kunlong Zhao$^1$, Jilu Jin$^2$, \\
		Xueqin Luo$^2$, Gongping Huang$^{1}$, Jingdong Chen$^2$, and Jacob Benesty$^3$
	\end{tabular} }

\address{$^1$School of Electronic Information, Wuhan University, 430072, Wuhan, China\\
	$^2$CIAIC, Northwestern Polytechnical University, Xi’an, Shaanxi, China\\
	$^3$INRS-EMT, University of Quebec, Montreal, Canada
	\vspace{-1.2em}}
\begin{document}
\ninept

\maketitle
\begin{abstract}
Differential microphone arrays offer a promising solution for far-field acoustic signal acquisition due to their high spatial directivity and compact array structure. A key challenge lies in designing differential beamformers that are continuously steerable and capable of enhancing target signals arriving from arbitrary directions. This paper studies the design of differential beamformers for circular arrays and proposes a novel framework that incorporates directional derivative constraints. By constraining the first-order derivatives of the beampattern at the desired steering direction to zero and assigning suitable values to higher-order derivatives, the beamformer is ensured to achieve its maximum response in the target direction and provide sufficient beam steering. This approach not only improves steering flexibility but also enables a more intuitive and robust beampattern design. Simulation results demonstrate that the proposed method produces continuously steerable beampatterns.
\end{abstract}

\section{Introduction}
\label{Sect-Intro}

Beamforming with microphone arrays plays a crucial role in speech and acoustic signal processing applications such as teleconferencing, hearing aids, and human–machine interaction~\cite{elko2008microphone, dmochowski2010microphone, gannot2004analysis, elko2000superdirectional, huang2025advances}. Among various beamforming techniques, differential beamforming has gained considerable attention owing to its compact implementation and inherently high directivity~\cite{elko2004differential, abhayapala2010higher, enzo2012on, huang2022fundamental, kolundzija2011spatiotemporal}.
Differential microphone array (DMA) design is typically achieved by enforcing a unit gain in the desired direction while placing nulls in unwanted directions, formulated in the short-time Fourier transform (STFT) domain~\cite{benesty2012study, benesty2023microphone, bernardini2017wave}.
Linear DMAs suffer from limited steering capabilities, typically exhibiting optimal performance only in the endfire direction~\cite{benesty2012study, jin2021steering}. To enable full azimuthal steering, linear superarrays incorporating both omnidirectional and bidirectional microphones~\cite{luo2023design, luo2024design}, as well as circular DMAs~\cite{benesty2015design, lovatello2018teerable, zhao2021design, wu2016directivity, buchris2020joint}, and other planar arrays have been introduced~\cite{bernardini2017efficient, zhao2021design, huang2018insights, wu2016directivity}. A common design strategy for circular DMAs, inherited from linear DMAs, is to enforce a unit gain in the target direction while placing nulls at undesired directions. However, this approach often leads to undesirable beampatterns and inaccurate steering behavior~\cite{huang2020continuously}.

Previous studies have attempted to address this issue by introducing symmetric constraints based on the geometry of uniform circular arrays (UCAs)~\cite{benesty2015design}. However, these constraints typically limit steering to a small set of fixed directions. Other approaches, such as the series expansion method~\cite{huang2017design, yan2015optimal}, offer full steering flexibility but rely on prior knowledge of the complete desired directivity pattern, which is often unavailable in practice. 
An alternative line of work extends the null-constrained design by incorporating the inherent symmetry of UCA configurations, enabling fully steerable beamformers~\cite{huang2020continuously, wang2021robust}. Nevertheless, this approach lacks theoretical rigor and intuitive interpretability, as it focuses solely on enforcing symmetric null placements without providing a clear justification for the underlying symmetry constraints.

To overcome these limitations, this paper introduces a novel formulation based on directional derivative constraints. By constraining the first-order derivatives of the beampattern at the desired steering direction to zero and assigning suitable values to higher-order derivatives, the beamformer is ensured to achieve its maximum response in the target direction and provide sufficient beam steering. This constraint offers a straightforward and intuitive means for achieving continuously steerable and accurately aligned beams toward the desired source. Simulation results support the theoretical analysis and confirm the effectiveness and robustness of the proposed method.

\section{Signal Model, Problem Formulation, and Performance Measures}
\label{Sect-SM-PB}

Consider a UCA of radius $r$, consisting of $M$ omnidirectional microphones. The array center is located at the origin of the Cartesian coordinate system, with azimuth angles measured counterclockwise from the $x$-axis ($\theta = 0$). Microphone~1 is placed on the $x$-axis. A farfield plane wave impinges on the array from the azimuth angle $\theta$ in an anechoic environment, propagating at the speed of sound $c = 340$~m/s.
The position of the $m$th microphone is given by $\mathbf{r}_{m} = \left[ r \cos\psi_m ~~~ r \sin\psi_m \right]^T$, where $\psi_{m} = 2\pi (m - 1)/M$, $m = 1, 2, \ldots, M$. The steering vector of length $M$ corresponding to the microphone arrays is written as~\cite{van2002optimum}
\begin{align}
	\label{d-vect}
	\mathbf{d}_{\theta} \left(\omega \right) &= \left[
	e^{\displaystyle \jmath \varpi \cos\left( \theta - \psi_1 \right) } ~~ e^{\displaystyle \jmath \varpi \cos\left( \theta - \psi_2 \right) } \right. \nonumber \\
	&  \left. \begin{array}{ccc} ~~~~~~~~~~~~~~~~~~~~~~~~~~  
		& \cdots  &  e^{\displaystyle \jmath \varpi \cos\left( \theta - \psi_M \right) } \end{array} \right]^T,
\end{align}
where $\jmath$ is the imaginary unit, $\varpi = \omega r/c$, $\omega=2 \pi f$ is the angular frequency, and $f>0$ is the temporal frequency.

Consider a desired signal $X(f)$ propagating from the angle $\theta_{\mathrm{s}}$, the frequency-domain observed signal can be written in a vector of length $M$ as
\begin{align}
	\label{Y-vect}
	\mathbf{y}\left( \omega \right) &=  \left[ \begin{array}{cccc}
		Y_1\left( \omega \right) & Y_2\left( \omega \right) & \cdots &
		Y_M\left( \omega \right) \end{array} \right]^T \nonumber \\
	&= \mathbf{d}_{\theta_{\mathrm{s}}} \left(\omega \right) X\left( \omega \right) +
	\mathbf{v}\left( \omega \right),
\end{align}
where $\mathbf{v}\left(\omega \right)$ is the zero-mean additive noise signal vector. 
The beamforming process can be presented as 
\begin{align}
	\label{z-out}
	Z\left( \omega \right) &=
	\mathbf{h}^H\left(\omega \right) \mathbf{y}\left( \omega \right) \nonumber \\
	&= \mathbf{h}^H\left(\omega \right) \mathbf{d}_{\theta_{\mathrm{s}}}\left(\omega \right) 
	X\left( \omega \right)
	+ \mathbf{h}^H\left(\omega \right) \mathbf{v}\left( \omega \right),
\end{align}
where the superscript $^H$ is the conjugate-transpose operator and
$Z\left(\omega \right)$ is the estimate of the desired signal. 

The distortionless constraint ensures that the desired signal can pass through the look direction without distortion, i.e.,
\begin{align}
	\label{const-oo}
	\mathbf{h}^H\left(\omega \right) \mathbf{d}_{\theta_{\mathrm{s}}} \left(\omega \right) =1.
\end{align}

To evaluate beamforming performance, we adopt three standard metrics: beampattern, ${\cal B} \left[ \mathbf{h}\left(\omega \right),\theta \right]$, white noise gain (WNG), ${\cal W}\left[\mathbf{h}\left(\omega \right) \right]$, and directivity factor (DF), ${\cal D}\left[\mathbf{h}\left(\omega \right) \right]$, defined as follows:
\begin{align}
	\label{BP-UCA}
	{\cal B} \left[ \mathbf{h}\left(\omega \right),\theta \right] &=
	\mathbf{h}^H\left(\omega \right) \mathbf{d}_{\theta}\left(\omega \right),
\end{align}
\begin{align}
	\label{WNG-define}
	{\cal W}\left[\mathbf{h}\left(\omega \right) \right]
	&= \frac{\left| \mathbf{h}^H\left(\omega \right) \mathbf{d}_{\theta_{\mathrm{s}}}\left(\omega \right) \right|^2}
	{\mathbf{h}^H\left(\omega \right) \mathbf{h}\left(\omega \right)},
\end{align}
and
\begin{align}
	\label{DF-define}
	{\cal D}\left[\mathbf{h}\left(\omega \right) \right]
	&=\frac{\left| \mathbf{h}^H\left(\omega \right) \mathbf{d}_{\theta_{\mathrm{s}}}\left(\omega \right) \right|^2}
	{\mathbf{h}^H\left(\omega \right) \mathbf{\Gamma}_{\mathrm{d}}\left(\omega \right) \mathbf{h}\left(\omega \right)},
\end{align}
where $\mathbf{\Gamma}_{\mathrm{d}}\left(\omega \right)$ is the coherence matrix of the diffuse noise, whose $(i,j)$th element is $\mathrm{sinc} \left( \omega \delta_{ij}/c \right)$, with $\delta_{ij} = 2 r \left| \sin \left[\left( i-j \right) \pi /M \right] \right|$ being the distance between microphones $i$ and $j$.

The frequency-independent directivity pattern of an $N$th-order differential beamformer steered to $\theta_{\mathrm{s}}$ is given by
\begin{align}
	\label{beam-ideal}
	{\cal B}_{N,\theta_{\mathrm{s}}} \left(\theta \right)
	&= \sum_{n=0}^N a_{N,n} \cos^n\left(\theta-\theta_{\mathrm{s}} \right),
\end{align}
where $a_{N,n} \in \mathbb{R}$, $n = 0, 1, \ldots, N$ are coefficients satisfying $\sum_{n=0}^N a_{N,n}=1$, which define the specific shape of the beampattern.

As seen from (\ref{beam-ideal}), the directivity pattern is symmetric about $\theta_{\mathrm{s}}$, i.e.,
\begin{align}
	\label{beam-ideal-sym}
	{\cal B}_{N,\theta_{\mathrm{s}}} \left( \theta_{\mathrm{s}} + \theta' \right) &=
	{\cal B}_{N,\theta_{\mathrm{s}}} \left( \theta_{\mathrm{s}} -\theta' \right), \ \theta' \in [0,\pi].
\end{align}
Consequently, when designing a differential beamformer, the main beam should point toward $\theta_{\mathrm{s}}$, and the beampattern must be symmetric about the line $\theta_{\mathrm{s}} \leftrightarrow \theta_{\mathrm{s}} + \pi$ for $\theta \in [0, \pi]$, i.e., ${\cal B} \left[ \mathbf{h}\left(\omega \right), \theta_{\mathrm{s}} + \theta \right] = {\cal B} \left[ \mathbf{h}\left(\omega \right), \theta_{\mathrm{s}} -\theta \right]$.
Dividing the ideal beampattern into two hemispheres centered on the look direction, if a null at $\theta_{N,n}^{+}$ lies in the upper hemisphere (i.e., within $\left[\theta_{\mathrm{s}}, \theta_{\mathrm{s}} + \pi \right]$), then by symmetry, a corresponding null must also appear at $\theta_{N,n}^{-}$ in the lower hemisphere (i.e., within $\left[\theta_{\mathrm{s}} - \pi, \theta_{\mathrm{s}} \right]$), with $(\theta_{N,n}^{+}+\theta_{N,n}^{-} )/ 2=\theta_{\mathrm{s}}$.
As a result, the following symmetry constraints should be added to the beampattern~\cite{huang2020continuously}:
\begin{align}
	\label{const-oo-proposed}
	{\cal B} \left[ \mathbf{h}\left(\omega \right), \theta_{N,n}^{+} \right] &=
	\mathbf{d}^H_{\theta_{N,n}^{+}} \left(\omega \right)  \mathbf{h}\left(\omega \right)
	= 0,   \\
	{\cal B} \left[ \mathbf{h}\left(\omega \right), \theta_{N,n}^{-} \right] &=
	\mathbf{d}^H_{\theta_{N,n}^{-}} \left(\omega \right)  \mathbf{h}\left(\omega \right)
	= 0.
\end{align}
By combining the null constraints, symmetry constraints, and the distortionless constraint, a linear system of equations can be formed and solved to obtain the beamformer~\cite{huang2020continuously}.
Although beam steering can be achieved by imposing symmetric null constraints, this approach lacks theoretical clarity and intuitive interpretability. This paper presents a more effective design method through a theoretical analysis.

\section{Steerable Differential Beamformers via Directional Derivative Constraints}

\begin{figure}[tb!]
	\centerline{\includegraphics[width=74mm]{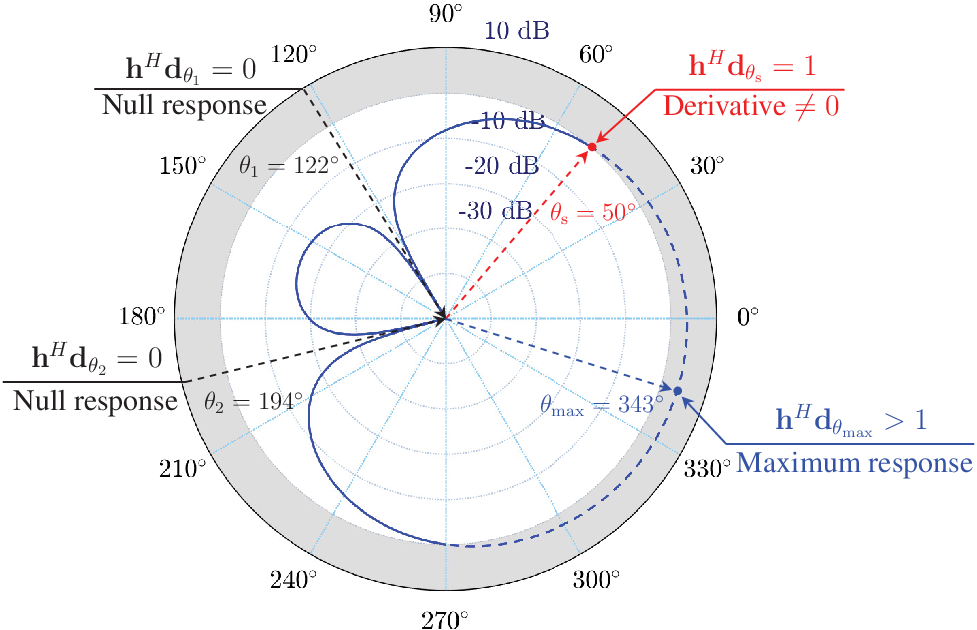}}\vskip -6pt
	\caption{Beampattern of the beamformer designed with null constraints at the specified null directions. The parameters are $M = 8$, a UCA with radius $r = 2.0$ cm, $f = 1$ kHz, $\theta_{\mathrm{s}} = 50^\circ$, and two nulls at $122^\circ$ and $194^\circ$.}
	\label{fig-BP-ula-uca}
\end{figure}

Generally, an $N$th-order differential beamformer is expected to have $N$ nulls. Following the conventional approach for linear DMAs, we start by enforcing unit gain in the desired direction and zero gain at the null directions.
To illustrate this, Fig.~\ref{fig-BP-ula-uca} shows examples using a UCA with eight sensors uniformly distributed on a circle of radius $r = 2.0$~cm. We aim to design a second-order hypercardioid beamformer with two nulls initially placed at $72^\circ$ and $144^\circ$ when the steering direction is $\theta_{\mathrm{s}} = 0^\circ$. When the beamformer is steered to $\theta_{\mathrm{s}} = 50^\circ$, the two nulls are correspondingly shifted to $122^\circ$ and $194^\circ$, where the null constraints are imposed.
As shown, although the beampatterns meet the imposed constraints, achieving unit gain in the target direction, i.e., $ 50^\circ$, and nulls at the specified locations, i.e., $122^\circ$ and $194^\circ$, they exhibit gains exceeding one in many other directions. This can lead to the amplification of noise, interference, or reverberation instead of suppression, which is clearly undesirable. The root cause lies in the inherent symmetry of the circular array, which is not properly accounted for in this beamforming approach.

Upon closer analysis, we identify a key limitation in the above design: constraining only the unit response at the desired direction and nulls at specific locations does not guarantee that the main lobe will be centered in the desired direction. Although the beamformer satisfies these constraints, its maximum response may deviate from the target direction, which clearly contradicts the original goal of beam steering.
Motivated by this insight, we propose an alternative constraint strategy. If we aim to ensure that the beamformer achieves its maximum response at the desired direction $\theta_{\mathrm{s}}$, then a more direct approach is to enforce that the first-order derivative of the beampattern with respect to direction equals zero at $\theta_{\mathrm{s}}$. This ensures that the desired direction corresponds to a local maximum, thereby guaranteeing correct beam steering.

We first consider the design of a first-order differential beamformer, with one null at $\theta_{1,n}$. The desired directivity pattern can be expressed as
\begin{align}
	\label{BP-2}
	{\cal B}_1 \left( \theta  \right)
	&= a_{1,0} + a_{1,1} \cos \left(\theta-\theta_{\mathrm{s}} \right).
\end{align}
It is easy to check that the first-order derivative of ${\cal B}_1 \left( \theta  \right)$
with respect to $\theta$ is
\begin{align}
	\label{BP-deriv-1st}
	{\cal B}_1^{(1)} \left( \theta  \right) =
	a_{1,1} \sin \left(\theta-\theta_{\mathrm{s}} \right).
\end{align}
We can easily check that  ${\cal B}_1^{(1)} \left(\theta_{\mathrm{s}} \right)=0$.
Consequently, the partial derivative of the beamformer's beampattern with respect to $\theta$ at $\theta_{\mathrm{s}}$ should also satisfy
\begin{align}
	\label{BP-deriv2}
	\left. \frac{\partial {\cal B} \left[ \mathbf{h} \left(\omega \right),\theta \right]}
	{\partial \theta } \right|_{\theta = \theta_{\mathrm{s}}} &=0.
\end{align}
The previous expression is equivalent to
\begin{align}
	\label{df-deriv3}
	\left[\mathbf{d}^H_{\theta_{\mathrm{s}}}\left(\omega \right) \right]^{(1)} \mathbf{h}\left(\omega \right)=0,
\end{align}
where $\left[\mathbf{d}_{\theta_{\mathrm{s}}}\left(\omega \right) \right]^{(1)}$
is the first-order partial derivative of $\mathbf{d}^H_{\theta} \left(\omega \right)$ at $\theta_{\mathrm{s}}$.

Consequently, the null constraints in a linear system of equations to design a first-order differential beamformer are
\begin{align}
	\label{lin-eqs-proposed}
	\mathbf{D}_{\mathrm{C}} \left( \omega \right) \mathbf{h}\left(\omega \right) 
	= \mathbf{i}_{1},
\end{align}
where
\begin{align}
	\label{D-matrix-2}
	\mathbf{D}_{\mathrm{C}} \left( \omega \right)
	\renewcommand\arraystretch{1.38} 
	= \left[ \begin{array}{c}
		\mathbf{d}^{H}_{\theta_{\mathrm{s}}}\left(\omega \right) \\
		\left[ \mathbf{d}^{H}_{\theta_{\mathrm{s}}}\left(\omega \right) \right]^{(1)} \\
		\mathbf{d}^H_{ \theta_{1,1}} \left(\omega \right) \\
	\end{array} \right] \in \mathbb{C}^{3 \times M}
\end{align}
is the constraint matrix that incorporates the distortionless constraint, null constraint, and directional derivative constraint, and $\mathbf{i}_{1} \in \mathbb{R}^{3 \times 1}$ is a unit vector.
To ensure clarity and avoid potential misinterpretation, we emphasize that constraining the first-order directional derivative of the beampattern to zero at the steering direction $\theta_{\mathrm{s}}$ is a necessary but not sufficient condition for a local maximum. The local maximum property at $\theta_{\mathrm{s}}$ is jointly guaranteed by the distortionless constraint (ensuring unit gain) and the constraints on the derivatives.

Similarly, for an $N$th-order differential beamformer, the beampattern is an $N$th-order function of $\theta$. To ensure that the maximum response occurs at the steering direction $\theta_{\mathrm{s}}$, the first-order derivatives of the beampattern with respect to $\theta$ should be zero at $\theta_{\mathrm{s}}$.
However, for higher-order ($>1$) derivatives, they can either be computed analytically or optimized under certain criteria. For example, if a desired beampattern is specified, the derivatives can be derived accordingly. Otherwise, all odd-order derivatives can be constrained to zero (which can be theoretically proved), while the second-order (or higher even-order) derivatives can be constrained to negative values, with their magnitudes influencing the beam shape (the resulting beam may no longer be symmetric). This work focuses on demonstrating that differential beamformers can be designed through derivative constraints, whereas the optimization of derivative values for beam shape control remains a topic for future research.
By combining the distortionless constraint, null constraints, and directional derivative constraints, we obtain the following linear system of equations:
\begin{align}
	\label{lin-eqs-proposed}
	\mathbf{D}_{\mathrm{C}} \left( \omega \right) \mathbf{h}\left(\omega \right) = \mathbf{i}_{\beta},
\end{align}
where
\begin{align}
	\label{D-matrix-2}
	\mathbf{D}_{\mathrm{C}} \left( \omega \right)
	\renewcommand\arraystretch{1.38} 
	= \left[ \begin{array}{c}
		\mathbf{d}^{H}_{\theta_{\mathrm{s}}}\left(\omega \right) \\
		\left[ \mathbf{d}^{H}_{\theta_{\mathrm{s}}}\left(\omega \right) \right]^{(1)}  \\
		\vdots \\
		\left[ \mathbf{d}^{H}_{\theta_{\mathrm{s}}}\left(\omega \right) \right]^{(N)}  \\
		\mathbf{d}^H_{\theta_{N,1} } \left(\omega \right) \\
		\vdots \\
		\mathbf{d}^H_{\theta_{N,N}} \left(\omega  \right)
	\end{array} \right]  \in \mathbb{C}^{(2N+1) \times M}
\end{align}
is the constraint matrix with
\begin{align}
	\label{df-derfine}
	\left[ \mathbf{d}^{H}_{\theta_{\mathrm{s}}}\left(\omega \right) \right]^{(q)}
	=  \left. \frac{\partial^q \mathbf{d}^H_{\theta } \left(\omega \right) }
	{\partial \theta^q} \right|_{\theta = \theta_{\mathrm{s}}},
\end{align}
for $q =1,2,\ldots,Q-1$, being the $q$th-order partial derivative of $\mathbf{d}^H_{\theta} \left(\omega \right)$ at $\theta_{\mathrm{s}}$, and $\mathbf{i}_{\beta} \in \mathbb{R}^{(2N+1) \times 1}$ is a vector.

In some cases, a null may have multiplicity greater than one, allowing additional constraints to be imposed by setting the derivatives at that null location to zero~\cite{chen2014design}. For example, if a null at $\theta_{N,n}$ has multiplicity $Q$, the corresponding constraints can be written as
\begin{align}
	\label{df-deriv}
	\left[ \mathbf{d}^{H}_{\theta_{N,n}} \left(\omega \right) \right]^{(q)}
	\mathbf{h}\left(\omega \right) &=0,
\end{align}
for $q =0,1,\ldots,Q-1$.

It can be verified that at least $2N + 1$ microphones are required to design a steerable $N$th-order differential beamformer. This is consistent with the conclusion in \cite{huang2017design}, which states that a minimum of $M = 2N + 1$ microphones is necessary for constructing a continuously steerable differential beamformer.
The simplest way to derive the beamformer is by maximizing the WNG, for which the solution is
\begin{align}
	\label{h-filter-MWNG-sym}
	\mathbf{h}_{\mathrm{MWNG}} \left(\omega \right)
	&= \mathbf{D}_{\mathrm{C}}^H\left(\omega \right)
	\left[ \mathbf{D}_{\mathrm{C}}\left(\omega \right) \mathbf{D}_{\mathrm{C}}^H\left(\omega \right) \right]^{-1} \mathbf{i}_{\beta}.
\end{align}

\vspace{2pt}
\section{Simulations}
\label{Sect-Simuls}

\begin{figure}[!t]
	\centerline{\includegraphics[width=81mm]{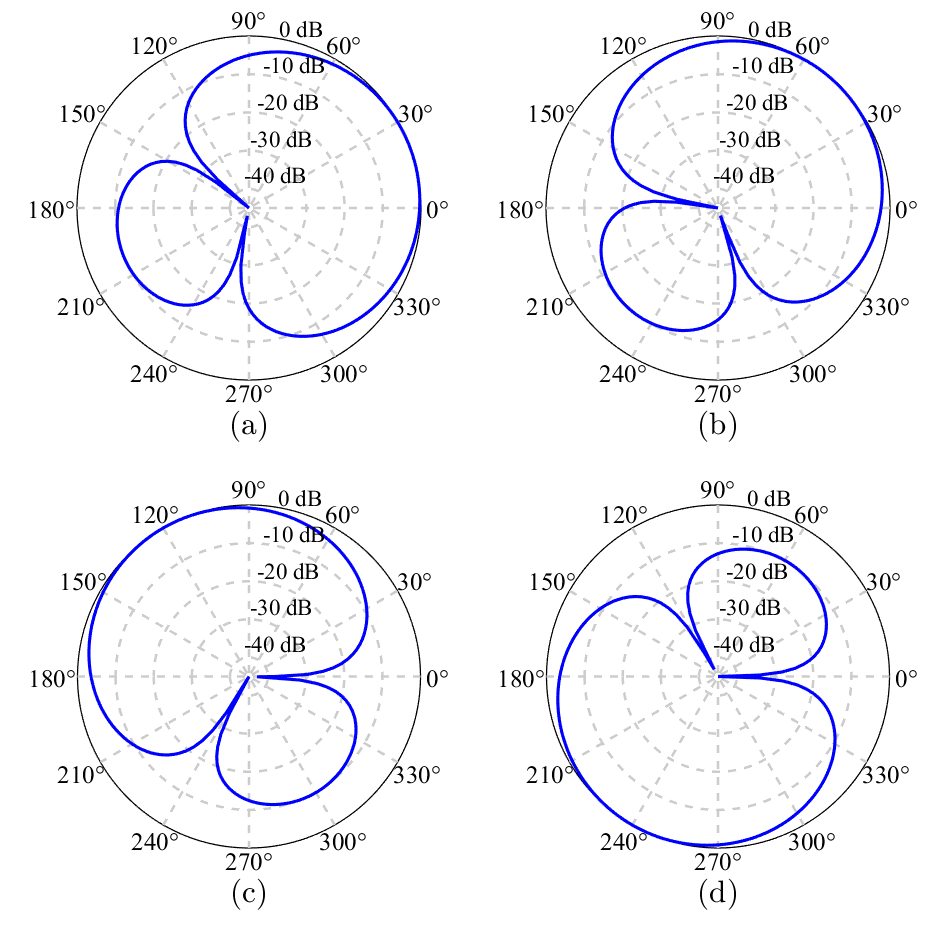}}
	\vskip -6pt
	\caption{Beampatterns at $f = 1$~kHz of first-order differential beamformers designed using the proposed method for steering directions:
		(a) $\theta_{\mathrm{s}} = 20^\circ$,
		(b) $\theta_{\mathrm{s}} = 50^\circ$,
		(c) $\theta_{\mathrm{s}} = 120^\circ$,
		and (d) $\theta_{\mathrm{s}} = 240^\circ$.}
	\label{fig-BP-1st}
\end{figure}

\begin{figure}[tb!]
	\centerline{\includegraphics[width=84mm]{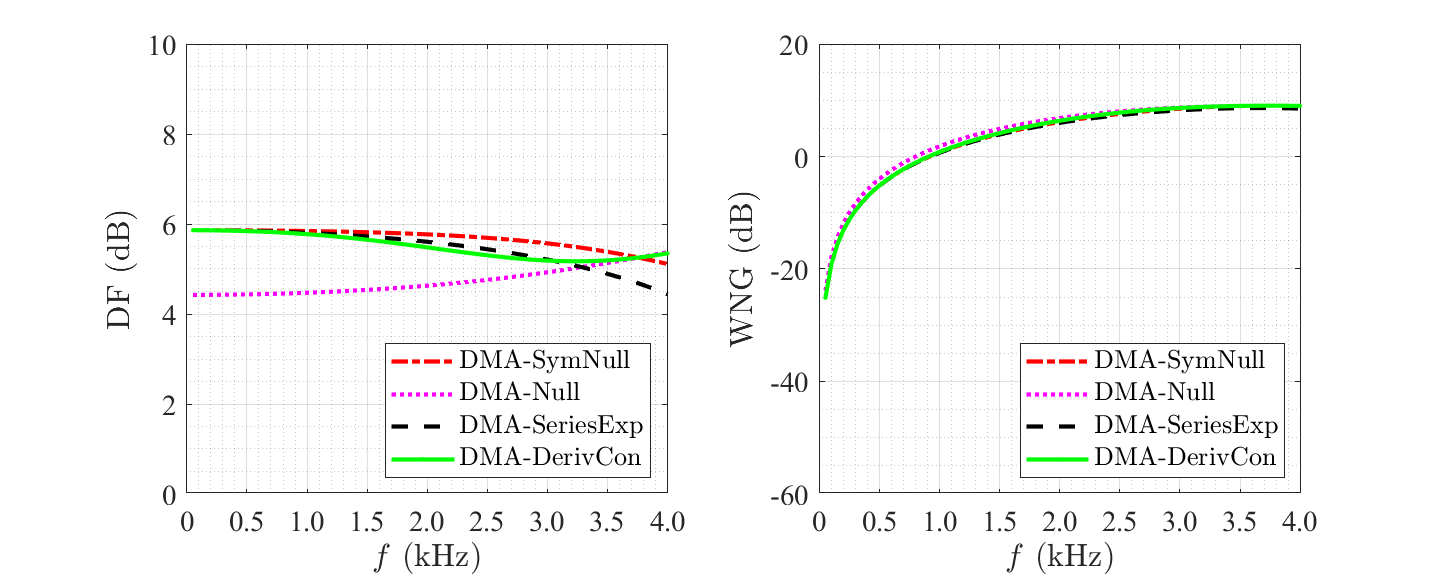}}\vskip -6pt
    \caption{DF and WNG comparison of first-order differential beamformers designed using the DMA-Null, DMA-SymNull, DMA-SeriesExp, and proposed DMA-DerivCon methods: (a) DF and (b) WNG. Conditions are $M = 8$, UCA with $r=2.0$~cm and $\theta_{\mathrm{s}} = 50^\circ$.}
	\label{fig-gain-1st}
\end{figure}

\begin{figure}[!t]
	\centerline{\includegraphics[width=81mm]{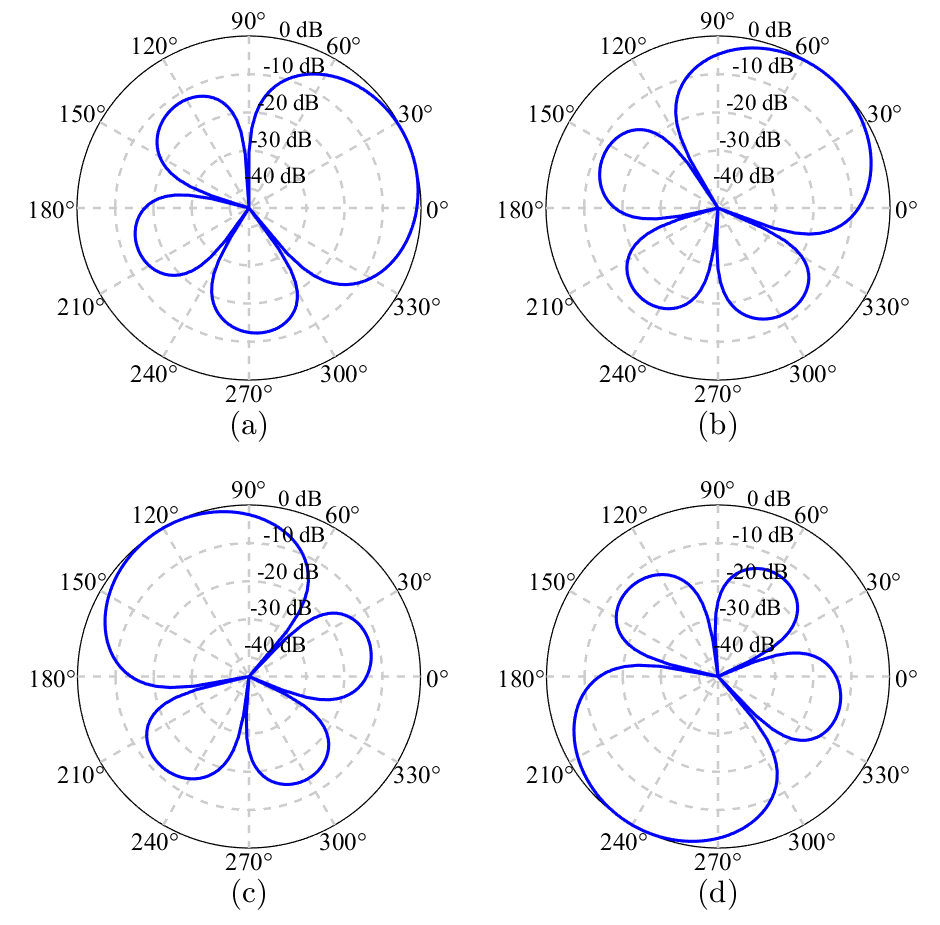}}
	\vskip -6pt
	\caption{Beampatterns at $f = 1$~kHz of second-order differential beamformers designed using the proposed method for steering directions:
		(a) $\theta_{\mathrm{s}} = 20^\circ$,
		(b) $\theta_{\mathrm{s}} = 50^\circ$,
		(c) $\theta_{\mathrm{s}} = 120^\circ$,
		and (d) $\theta_{\mathrm{s}} = 240^\circ$.}
	\label{fig-BP-2nd}
\end{figure}

\begin{figure}[tb!]
	\centerline{\includegraphics[width=84mm]{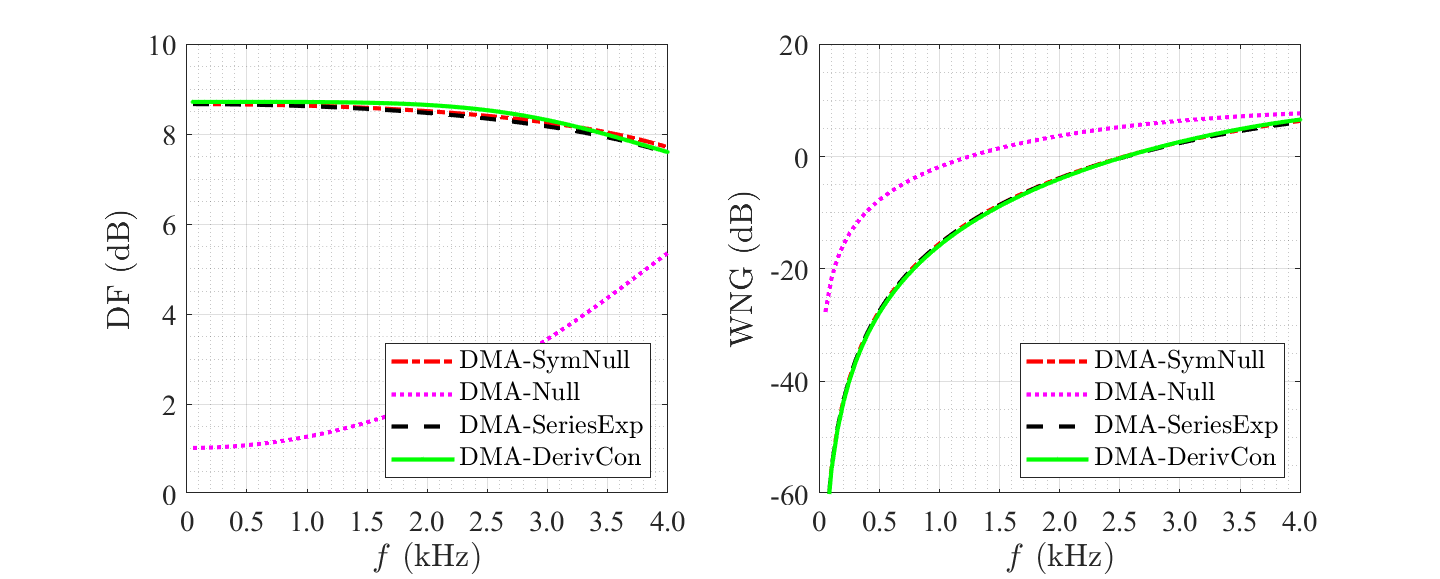}}\vskip -6pt
	\caption{DF and WNG comparison of second-order differential beamformers designed using the DMA-Null, DMA-SymNull, DMA-SeriesExp, and proposed DMA-DerivCon methods:
		(a) DF and (b) WNG. Conditions are $M = 8$, UCA with $r=2.0$~cm and $\theta_{\mathrm{s}} = 50^\circ$.}
	\label{fig-gain-2nd}
\end{figure}

In this section, we evaluate the performance of the proposed steerable differential beamformers (DMA-DerivCon) based on directional derivative constraints. We consider a UCA of radius $r=2.0$~cm, consisting of $8$ omnidirectional microphones.
For comparison, the conventional null-constrained method (DMA-Null)~\cite{benesty2012study}, the symmetric null-constrained method (DMA-SymNull)~\cite{huang2020continuously}, and the series expansion method (DMA-SeriesExp)~\cite{huang2017design} are also included.

In the first simulation, we design a first-order differential beamformer with a single null at $120^\circ$. When the desired steering direction is set to $\theta_{\mathrm{s}}$, the corresponding null is placed at $\theta_{\mathrm{s}} + 120^\circ$, with $\mathbf{i}_{\beta} = [1, ~0, ~0]^T$.
To demonstrate the steering flexibility of the proposed method, Fig.~\ref{fig-BP-1st} presents the designed beampatterns for four arbitrarily selected steering directions: $\theta_{\mathrm{s}} = 20^\circ$, $50^\circ$, $120^\circ$, and $240^\circ$. It can be observed that the main lobes are correctly steered toward the desired directions, while deep nulls are formed at the expected positions.
Fig.~\ref{fig-gain-1st} compares the DF and WNG of several first-order differential beamforming methods, including DMA-Null, DMA-SymNull DMA-SeriesExp, and the proposed DMA-DerivCon. As seen, the proposed DMA-DerivCon exhibits a more balanced DF across the entire frequency band while maintaining a WNG comparable to that of the other methods.

In the second simulation, we design a second-order differential beamformer with two nulls placed at $120^\circ$ and $240^\circ$. When the desired direction is set to $\theta_{\mathrm{s}}$, the corresponding nulls are located at $\theta_{\mathrm{s}} + 120^\circ$ and $\theta_{\mathrm{s}} + 240^\circ$, respectively, with $\mathbf{i}_{\beta} = [1, ~0, ~-2, ~0, ~0]^T$.
To demonstrate the beamformer's null-steering capability, Fig.~\ref{fig-BP-2nd} shows the resulting beampatterns for several selected steering directions. As shown, the main lobes are accurately steered to the desired directions, and deep nulls are formed at the expected interference angles. This validates the ability of the proposed method to impose multiple directional constraints while maintaining a distortionless response in the target direction.
Fig.~\ref{fig-gain-2nd} compares the DF and WNG of several second-order differential beamforming methods, including DMA-Null, DMA-SymNull, DMA-SeriesExp, and the proposed DMA-DerivCon. The proposed DMA-DerivCon maintains a relatively high and smooth DF across the frequency band while achieving a WNG comparable to that of the other methods. This demonstrates that the derivative-constrained design remains effective for higher-order differential beamforming without incurring a noticeable degradation in noise robustness.

\section{Conclusions}
\label{Sect-Conclu}

Designing steerable beamformers with UCAs is crucial for many sound capture applications. In this paper, a directional derivative-constrained method was proposed for the design of steerable differential beamformers with UCAs. It was demonstrated that constraining only the desired and null directions failed to guarantee correct steering, as the maximum response might not appear in the intended direction. To address this issue, the proposed method introduced directional derivative constraints by forcing the first-order derivatives of the beampattern at the desired steering direction to zero and assigning suitable values to higher-order derivatives. This ensured that the beamformer achieved its maximum response in the target direction and provided continuous steering capability. Simulation results validated the theoretical analysis and confirmed that the proposed beamformer could produce well-formed beampatterns with accurate steering. The method, therefore, offers a practical and effective solution for steerable differential beamforming in circular array configurations.

\newpage  

\end{document}